\documentclass[12pt]{article}
\usepackage{graphicx}
\usepackage{setspace}
\usepackage{array}

\usepackage{amsmath}
\usepackage{amssymb}
\usepackage{epstopdf}
\usepackage{authblk}
\usepackage{cite}
\usepackage{fullpage}
\usepackage{float}
\usepackage{color}
\usepackage{subfig}
\title{\bf Emergence of chaotic hysteresis in a second-order non-autonomous chaotic circuit \footnote{Preprint submitted to Chaos Solitons \& Fractals}}
\author[1]{G. Sivaganesh}
\affil[1]{\it Department of Physics, Alagappa Chettiar Government College of Engineering and Technology, Karaikudi, Tamilnadu-630 003, India}
\author[2]{K. Srinivasan}
\affil[2]{\it Department of Physics, Nehru Memorial College, Puthanampatti, Tiruchirapalli, Tamilnadu - 621 007, India (Affiliated to Bharathidasan University, Tiruchirapalli, Tamilnadu - 620 024, India)}
\author[3]{T. Fonzin Fozin}
\affil[3]{\it Department of Electrical and Electronic Engineering, Faculty of Engineering and Technology (FET), University of Buea, P.O. Box 63, Buea, Cameroon}
\author[4]{R. Gladwin Pradeep}
\affil[4]{\it Department of Physics, KCG College of Technology, Chennai - 600 097, India}
\date{\today}
\begin{document}
\maketitle
\begin{abstract}
The observation of hysteresis in the dynamics of a third-order autonomous chaotic system namely, the {\it{Chua's}} circuit has been reported recently \cite{Gomes2023}. In the present work, we make a detailed study on the emergence of dynamical hysteresis in a simple second-order non-autonomous chaotic system namely, the {\it{Murali-Lakshmanan-Chua }} (MLC) circuit. The experimental realization of chaotic hysteresis is further validated by numerical simulation and analytical solutions. The presence of chaotic hysteresis in a second-order non-autonomous electronic circuit is reported for the first time. Multistable regions are observed in the dynamics of MLC with constant bias.\\

\end{abstract}
\doublespace

\section{Introduction}

The rise of computers and computing power in the latter part of the 20th century facilitated the numerical study of many dynamical systems, reigniting interest in understanding chaotic dynamical systems. Chaotic behavior, which is prevalent in nature, is characterized by extreme sensitivity to initial conditions, where even a small variation in initial conditions can result in a vastly different future state \cite{Lorenz1963}. Understanding chaotic dynamics is crucial in comprehending a wide range of natural systems. Furthermore, artificial dynamical systems were created as part of the effort to grasp chaotic dynamics, which led to the discovery of phenomena such as stochastic resonance, chaotic synchronization, and chaos control \cite{Lichtenberg1983,Anishchenko1992,Pecora1998,Pikovsky2003}. Chaotic electronic circuits are important for studying chaos because they are easy to construct and control \cite{Lakshmanan1995a,Lakshmanan2003}. {\it{Chua's}} circuit is one of the most extensively studied chaotic circuits and has led to the experimental realization of stochastic resonance, chaotic synchronization, and control of chaos \cite{Matsumoto1984,Matsumoto1985,Afraimovich1986,Srinivasan2023}. It is a piece-wise linear third-order autonomous system that has also been studied analytically \cite{Sivaganesh2022}. Recent research has reported hysteresis behavior in {\it{Chua's}} circuit with a DC offset voltage \cite{Gomes2023}. The variation of the DC offset from negative to positive values yields different dynamical paths, resulting in hysteresis in the system dynamics. The Murali-Lakshmanan-Chua (MLC) circuit, a second-order non-autonomous circuit constructed with {\it{Chua's}} diode as the nonlinearity \cite{Kennedy1992}, has also been instrumental in the experimental realization of various chaotic dynamics \cite{Murali1994a,Lakshmanan1995}. The MLC circuit and its variants have wide applications in secure communication, image encryption, electronic noise generation, chaotic logic gates, and reservoir computing \cite{Lakshmanan1997,Ugur2005,Murali2009,Murali2018,Zhiqiang2022}. Furthermore, analytical solutions for a class of  second-order chaotic systems have been reported recently \cite{Sivaganesh2023}.

In this study, we discuss the appearance of chaotic hysteresis in the behavior of the MLC circuit resulting from the introduction of a DC offset voltage in the circuit.  We examine the variations of the offset voltage that lead to the hysteresis phenomenon in the system dynamics and validate our electronic circuit experimental results through numerical and analytical results. The paper is structured as follows: in section \ref{exp_sec}, we present the hysteresis phenomenon observed in the MLC circuit using experimental phase portraits; in section \ref{num_sec}, we present the numerical simulation results that explain the emergence of hysteresis through bifurcation diagrams and the multistability in the system is analyzed using offset boosting method; and in section \ref{ana_sec}, we report the analytical solutions developed for the normalized circuit equations to study the hysteresis phenomenon. Finally, we provide our conclusions in section \ref{sec_con}.

\section{Chaotic hysteresis: Experimental results}
\label{exp_sec}

\begin{figure}
\begin{center}
\includegraphics[scale=1.0]{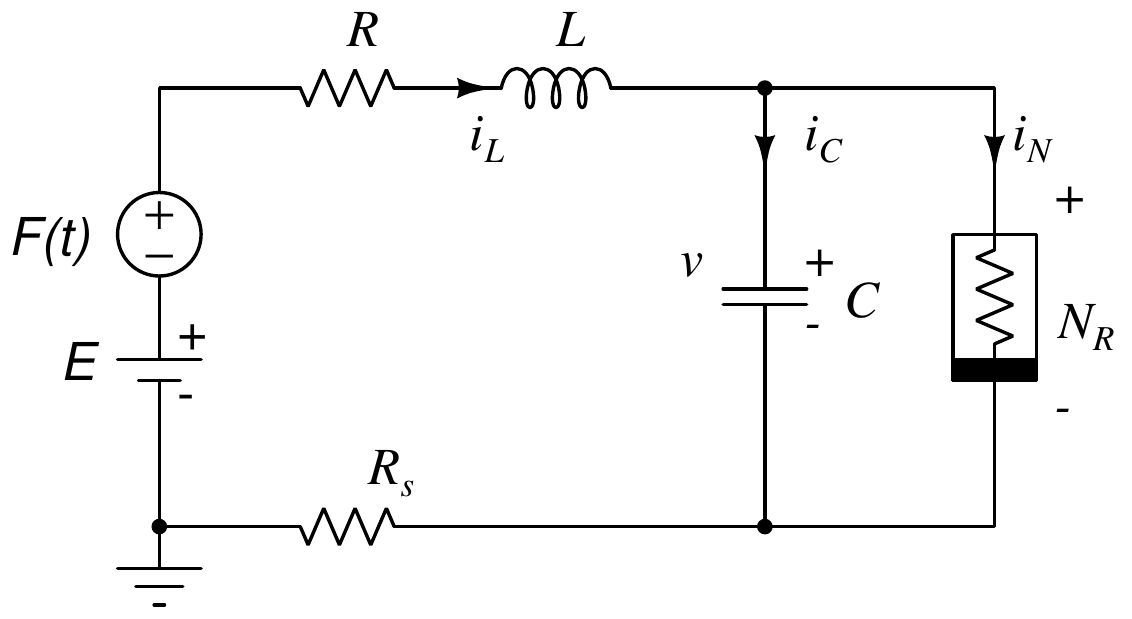}
\caption{Circuit realization of periodically driven MLC circuit with a DC offset voltage $E$.  Here, $N_R$ is the {\it{Chua's}} diode. The parameter values of the other elements are fixed as $L=18.0~mH$, $C=10.0~nF$, $R=1340~\Omega$ and $R_s=20~\Omega$.}
\label{mlc_cir}
\end{center}
\end{figure}

The circuit realization of the simple non-autonomous (MLC) circuit is shown in Fig.~\ref{mlc_cir}.  It contains a capacitor, an inductor, a linear resister, an external periodic forcing and only one nonlinear element, namely, the Chua's diode $(N_R)$. In order to  measure the inductor current $i_L$ in our experiments, we insert a small  current sensing resister $R_s$ as shown Fig. \ref{mlc_cir}. 

\begin{figure}
\begin{center}
\includegraphics[scale=0.66]{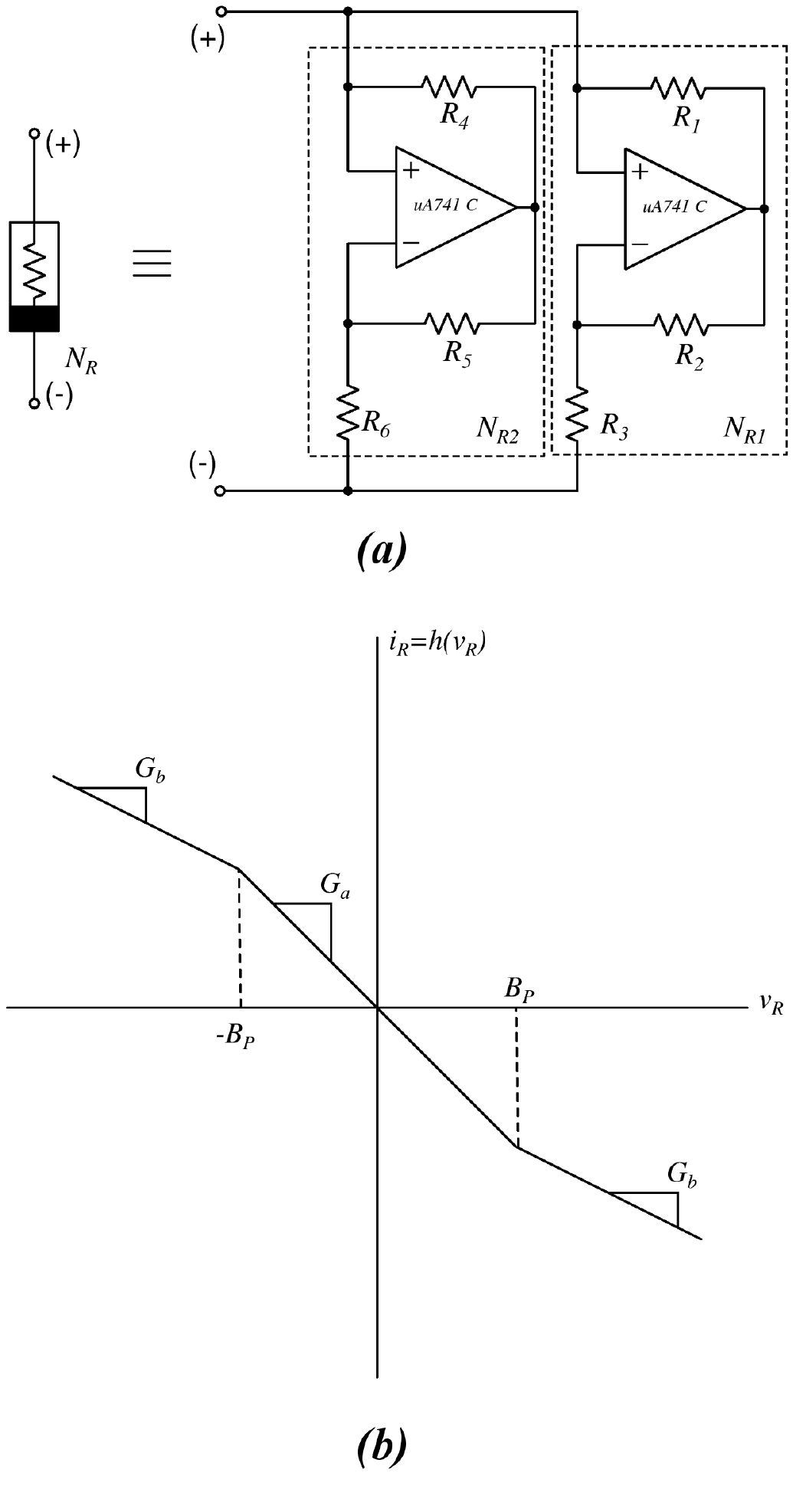}
\caption{a) Schematic realization of {\it{Chua's}} diode $N_R$ using Op-Amps. (b) Voltage-Current characteristics of {\it{Chua's}} diode indicating its piecewise-linear nature.}
\label{chua_diode}
\end{center}
\end{figure}
By applying Kirchhoff's laws to this circuit, the governing equations for  the voltage $v$ across the capacitor $C$ and the current $i_L$ through the inductor $L$ are represented by the following set of two first-order non-autonomous differential equations as
\begin{subequations}
\begin{eqnarray}
C {dv \over d\tau }  &=&  i_L - g(v), \\
L {di_L \over d\tau }  &=&  -(R + R_s) i_L - v + F sin( \Omega \tau) + E,
\end{eqnarray}
\label{exp_1}
\end{subequations}
where $F(t)$ is the sinusoidal periodic force with amplitude $F$ and angular frequency $\Omega$. The term $E$ represents a constant voltage added in addition to the sinusoidal voltage source ($F(t)$) is called as the $DC$ offset voltage which can be negative or positive. The value of $DC$ offset voltage shifts the signals to the same amount of its value. The term $g(v)$ represents the $v - i$ characteristics of the {\emph{Chua's}} diode as shown in Fig.~\ref{chua_diode} and is given by
\begin{subequations}
\begin{eqnarray}
g(v) = G_b v + 0.5(G_a - G_b)[|v+B_p|-|v-B_p|],
\end{eqnarray}
or in the piecewise-linear form it is given as
\begin{eqnarray}
g(v) =
\begin{cases}
{G_b}v+({G_a}-{G_b}) & \text{if $v \ge 1$}\\
{G_a}v & \text{if $|v|\le 1$}\\
{G_b}v-({G_a}-{G_b}) & \text{if $v \le -1$}
\end{cases}
\end{eqnarray}
\label{exp_2}
\end{subequations}
\begin{figure}
\begin{center}
\includegraphics[scale=0.5]{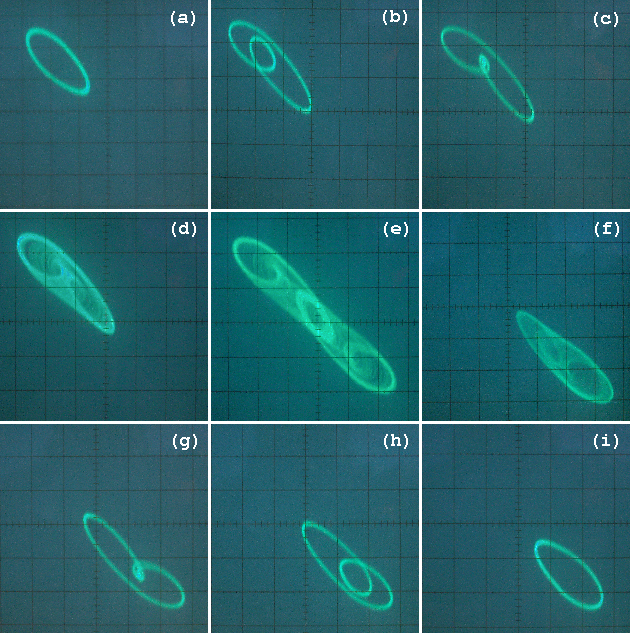}
\caption{Chaotic hysteresis observed in MLC circuit with variation of offset voltage $E$. The system is set for a one-band chaotic attractor state at $F = 0.11~V$ with $E=0$. Phase-portraits in the left-half of $v-i_L$ plane: (a) $E=-0.02~V$, period-1T LC (b) $E=-15~mV$, period-2T LC (c) $E=-8.0~mV$, period-4T LC (d) $E=-2.5~mV$, one-band chaos and (e) $E=0.1~mV$, double-band chaotic attractor. Phase-portraits in the right-half of $v-i_L$ plane: (f) $E=2.5~mV$, one-band chaos (g) $E=8.0~mV$, period-4 T LC (h) $E=15~mV$, period-2T LC (i) $E=0.02~V$, period-1T LC. LC-Limit Cycle.}
\label{exp_fig}
\end{center}
\end{figure}
The construction of {\it{Chua's}} diode using operational amplifiers (Op-Amps) and its corresponding {\it{Voltage-Current}} characteristics is shown in Fig. \ref{chua_diode}. The parameters of the circuit  elements are fixed at $L=18.0~mH, C=10.0~nF, R=1340~\Omega, R_s = 20~\Omega, G_a=-0.76~mS, G_b=-0.41~mS$ and $B_p=1.0~V$ and the frequency $(\nu = \Omega/2\pi)$ of the external forcing  source is $8890~Hz$. The parameters $F(t)$ and $E$ are taken as control parameters. The dynamical state of the circuit can be kept either in the chaotic or periodic state and the offset voltage $E$ is varied to observe the changes in the dynamics of the system.  
Figure \ref{exp_fig} shows the experimental dynamics of the $MLC$ circuit by varying the offset voltage $E$ with the circuit initially operated at the one-band chaotic attractor state for $F=0.11~V$ when $E=0$. Hence, in the absence of the DC offset, the system operates in the one-band chaotic attractor state. The DC offset voltage is increased from negative to positive values and the dynamics observed in the $v-i_L$ plane is reported. A period-1 limit cycle is observed for $E=-0.02~V$ leading to period-2 and period-4 limit cycles for $E=-15~mV$ and $E=-8.0~mV$ as shown in Fig. \ref{exp_fig}(a), \ref{exp_fig}(b) and \ref{exp_fig}(c), respectively. Further increase in $E$ leads to an one-band chaotic attractor in the left-half plane for $E=-2.5~mV$ followed by a double-band chaotic attractor at $E=0.1~mV$ as shown in Fig. \ref{exp_fig}(d) and \ref{exp_fig}(e), respectively. With increase in $E$, the dynamics of the system jumps to the right-half plane and results in the evolution of an one-band chaotic attractor shown in Fig. \ref{exp_fig}(g) at $E=2.5~mV$. Further increase in $E$ indicates a reverse period-doubling sequence represented by the period-4, period-2 and period-1 limit cycles for the values $E=8.0~mV$, $E=15~mV$ and $E=0.02~V$ as shown in Fig.  \ref{exp_fig}(g), \ref{exp_fig}(h) and \ref{exp_fig}(i), respectively. However, when we start from the period-1 limit cycle at the right-half plane, the decrease in $E$ results in the system dynamics indicated by Figs. \ref{exp_fig}(i), \ref{exp_fig}(h) and \ref{exp_fig}(g), \ref{exp_fig}(f) and gives rise to the double-band chaotic attractor of Fig. \ref{exp_fig}(e). Further decrease in $E$ results in the jumping of the dynamics to the left-half plane through the evolution of the one-band chaotic attractor shown in Fig. \ref{exp_fig}(d) followed by the reverse period-doubling sequence of the periodic attractors in the left-half plane. The region of $E$ encompassing the evolution of one-band chaotic attractor in the left-half plane followed by the double-band chaotic attractor leading to the one-band chaotic attractor in the right-half plane when $E$ is increased from negative values and the similar dynamics observed when $E$ is decreased such that the one-band chaotic attractor transits from the right-half plane to the left-half plane through the evolution of the double-band chaotic attractor represents the region of chaotic hysteresis observed in the circuit system.

\section{Chaotic hysteresis: Numerical results}
\label{num_sec}

In this section, we report the emergence of chaotic hysteresis in the MLC circuit through numerical simulation of the normalized circuit equations.

\subsection{Hysteresis evidence}
\label{num_sec1}

The normalized equations obtained from Eq. (\ref{exp_1}) is written as
\begin{subequations}
\begin{eqnarray}
\dot x  &=&  y - g(x), \\ 
\dot y  &=&  -\sigma y - \beta x + f sin(\omega t)+c,~~~~~\left(\cdot = \frac{d}{dt}\right)
\end{eqnarray}
\label{num_1}
\end{subequations}
where $t = (\tau G/C)$, $\sigma = (\beta + \nu \beta)$, $\beta = (C/LG^2)$, $ \nu = GR_s$, $ a  = G_a/G$,  $ b = G_b/G$, $f = (F \beta/B_p)$, $\omega = (\Omega C/G)$, $c = (E \beta/B_p)$, $ G = 1/R$. The nonlinear function $g(x)$ is written in its normalized form as
\begin{subequations}
\begin{eqnarray}
g(x) = bx + 0.5(b - a)[|x+1|-|x-1|]
\end{eqnarray}
In piecewise-linear form $g(x)$ is
\begin{eqnarray}
g(x) =
\begin{cases}
bx+(a-b) & \text{if $x \ge 1$}\\
ax & \text{if $|x|\le 1$}\\
bx-(a-b) & \text{if $x \le -1$}
\end{cases}
\end{eqnarray}
\label{num_2}
\end{subequations}

\begin{figure}
\begin{center}
\includegraphics[scale=0.23]{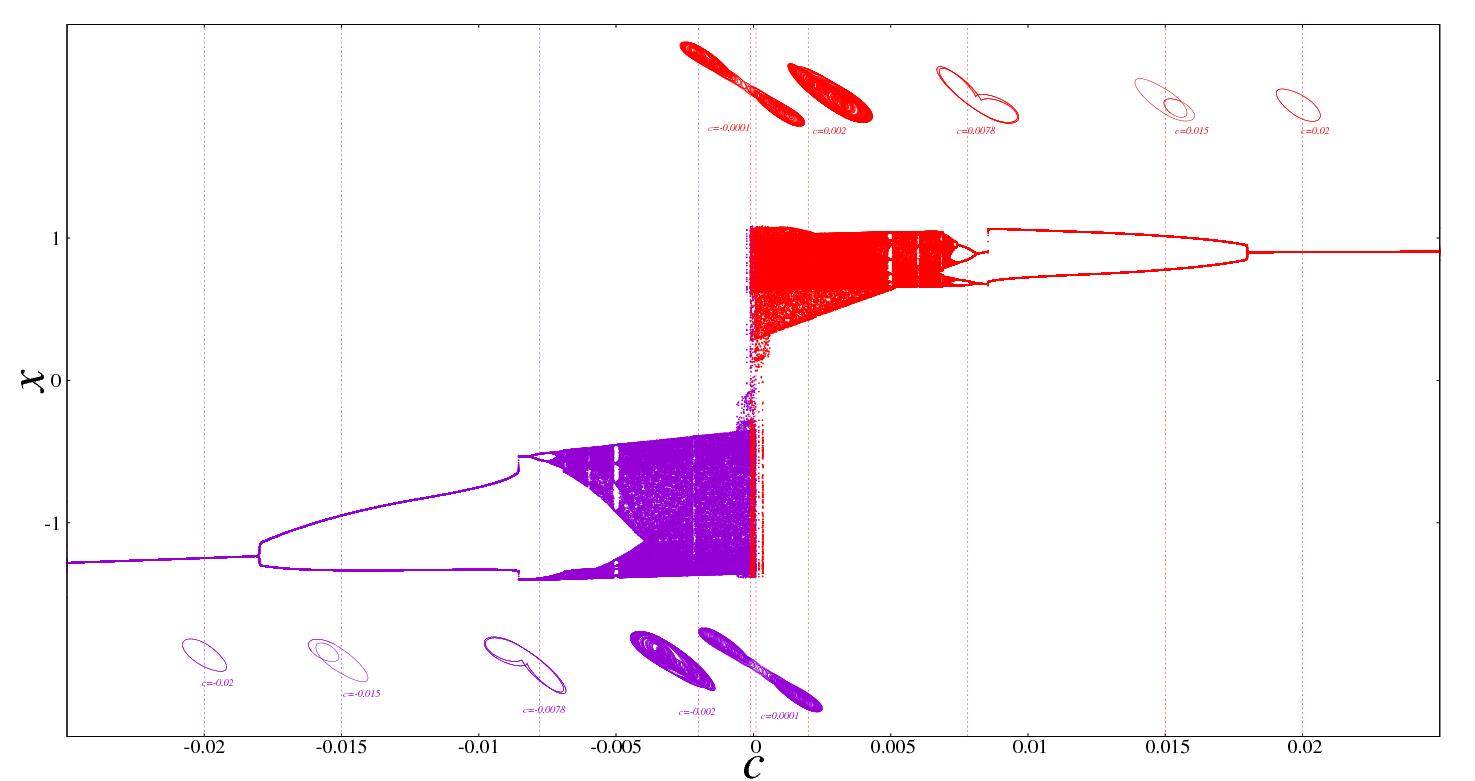}
\caption{Numerical results: One-parameter bifurcation diagram in the $c-x$ plane obtained in the range $-0.025 <c < 0.002$ for the initial conditions ($x_0, y_0$) =($-0.1, -0.1$) is indicated in violet while the bifurcation diagram in the range $0.025 > c > -0.002$ indicated in red is obtained for the initial condition $(x_0, y_0) = (0.1, -0.1)$. The region of the bifurcation diagram in violet corresponds to the attractor in the LHP while the region in red corresponds to the attractors in the RHP of the phase space. The vertical dashed lines in violet/red indicate the dynamics of the system at that point of $c$ which is also represented by the corresponding colored phase-portraits in the $(x-y)$ plane. The system jumps from the LHP to RHP at $c=0.002$ in the one-band chaotic state while the dynamics jumps from the RHP to LHP at $c=-0.002$ in the one-band chaotic attractor state. The region of $c$ in the range $-0.002 \le c \le 0.002$ indicates the region of chaotic hysteresis.}
\label{num_fig1}
\end{center}
\end{figure}

\begin{figure}
\begin{center}
\includegraphics[scale=0.5]{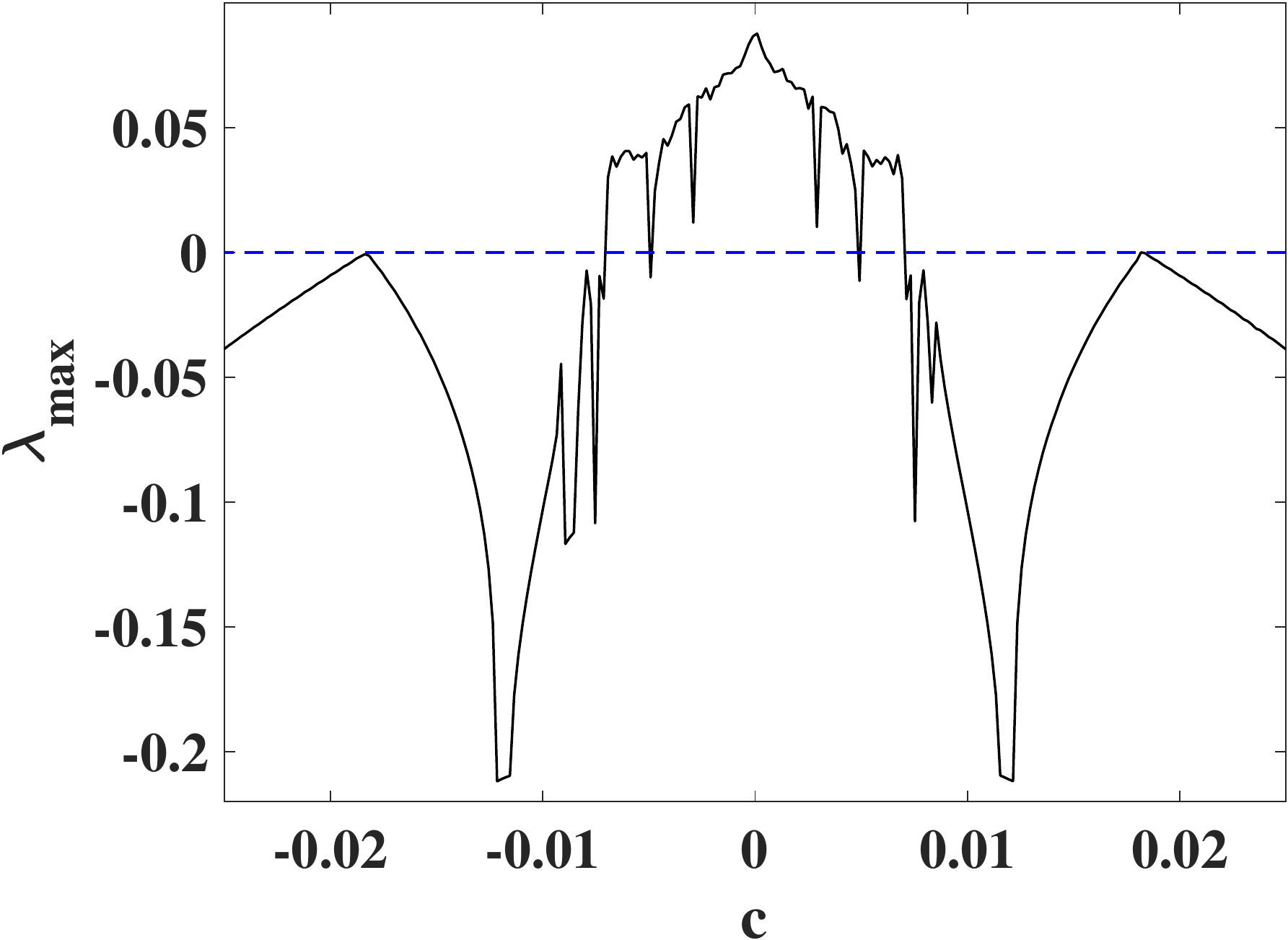}
\caption{Numerical result: Maximum Lyapunov exponent when varying parameter $c\in [-2.5,2.5]\times 10^{-2}$ in the MLC oscillator.}
\label{num_fig2}
\end{center}
\end{figure}

The phenomenon of chaotic hysteresis is studied using the one-parameter bifurcation diagram obtained through numerical simulation of the circuit equations. The normalized system parameters are fixed as $a=-1.02, b=-0.55, \beta=1.0, \nu=0.015, \omega=0.75$. Figure \ref{num_fig1} shows the one-parameter bifurcation diagram obtained for two different initial conditions and for the existence of the attractors in the left-half plane and right-half plane. The system is initially set at the one-band chaotic attractor state for $f =0.1104$ when the offset voltage parameter $c=0$. The period-doubling sequence leading to chaos obtained for the initial condition $(x_0, y_0) = (-0.1, -0.1)$ by sweeping the parameter $c$ in the range $-0.025 < c < 0.002$ is represented in violet while that obtained for the initial condition $(x_0, y_0) = (0.1, -0.1)$ by sweeping the parameter $c$ in the range $0.025 > c > -0.002$ is represented in red. The periodic and the chaotic attractors plotted at their respective positions are indicated by the vertical lines of their respective colors. The values of $c$ in the range $-0.025< c < 0.002$ represented in violet indicate the attractors in the left-half plane (LHP) and the attractors in the right-half plane (RHP) are indicated in red in the range $0.025 > c > -0.002$. In particular, the system exhibits chaotic attractors in the LHP at the values $c=-0.002$ and $c=0$ and a double-band chaotic attractor evolves at $c=0.0001$. At $c=0.002$, the system jumps to the RHP as a one-band chaotic attractor. Similarly, varying $c$ in the range $0.025 c  -0.002$ represented in red indicates the existence of attractors in the RHP with one-band and double-band chaotic attractors observed at the values $c=0.002$ and $c=-0.0001$, respectively. At $c=-0.002$, the system jumps to the LHP as a one-band chaotic attractor. Hence, the parameter range $-0.002 \ge c \ge 0.002$ indicates the emergence and existence of chaotic hysteresis when it is initially operated in the one-band chaotic attractor state. It has to be noted that the system exists in the one-band chaotic attractor state for $c=0$ and its location in the phase plane corresponds to the dynamical path of operation i.e. either from the LHP or RHP. The maximum Lyapunov exponent of the MLC circuit obtained as a function of $c$ indicating the regions of periodic and chaotic behavior is as shown in Fig. \ref{num_fig2}.

\section{Chaotic hysteresis: Analytical solutions}
\label{ana_sec}


In this section, we report the explicit analytical solutions developed for the state equations of the MLC circuit with offset voltage and study the hysteresis of the system dynamics using the analytical solutions. The MLC circuit system represented by Eq. (\ref{num_1}) is solved for the variables $x(t), y(t)$ in each of the piecewise-linear region and the solutions are matched across the break-point regions for generating the attractors in the phase plane. The analytical solutions for each region is summarized as follows.\\


\subsection{$D_0$ region}


In this region, $g(x) = ax$ and the state equations are written as
\begin{subequations}
\begin{eqnarray}
{dx \over dt }  &=&  y - a x, \\
{dy \over dt }  &=&  -\beta y - \nu \beta y -\beta x + f \sin( \omega t) + c,
\end{eqnarray}
\label{ana1}
\end{subequations}
The Jacobian matrix determining the stability of the fixed point $p_0 = (-\frac{c}{a+\beta},\frac{ac}{2 \beta+\nu \beta})$ in this region is 
\begin{equation}
J_0 =
\begin{pmatrix}
-a &&& 1 \\
-\beta &&& - \beta - \nu \beta \\
\end{pmatrix},
\label{ana2}
\end{equation}
and the fixed point $p_0$ is a {\it{saddle}} or {\it{hyperbolic fixed point}}. The fixed point is $(0,0)$ when $c=0$ and its location in the phase-plane changes with the variation of $c$.
Equation (\ref{ana1}) written as a second-order differential equation is given as
\begin{equation}
{\ddot y} + {A_1 \dot y} + B_1 y = \Delta_1 + a f \sin(\omega t) + f \omega \cos(\omega t),
\label{ana3}
\end{equation}
where $A_1 = \beta + \nu \beta + a$, $B_1=\beta+a (\beta+ \nu \beta)$ and $\Delta_1 = a c$. For the given values of the system parameters, the roots of Eq. (\ref{ana3}) $m_{1,2} = \frac{-A_1}{2} \pm \frac{\sqrt{{A_1}^2-4B_1}}{2}$ are real and distinct and the state variables are obtained as
\begin{subequations}
\begin{eqnarray}
y(t) &=& C_1 e^ {m_1 t} + C_2 e^ {m_2 t} +E_1 +E_2 \sin (\omega t) + E_3 \cos (\omega t), \\
x(t) &=& \frac{1}{\beta}(-(\beta + \nu \beta) y - \dot{y} + c + f \sin (\omega t)),
\end{eqnarray}
\label{ana4}
\end{subequations}
where, the constants $E_1, E_2, E_3,C_1,C_2$ are
\begin{subequations}
\begin{eqnarray}
E_1 =&& \frac{\Delta_1}{B_1} \\
E_2  =&&  \frac {f  {\omega} ^2 (A_1-a) + a f B_1}{A_1^2 {\omega} ^2 + (B_1-{\omega} ^2)^2}  \\
E_3  =&&  - \frac {f  \omega (A_1 a +\omega ^2 -B_1)}{A_1^2 {\omega} ^2 + (B_1-{\omega} ^2)^2}\\
C_1 =  &&\frac{e^ {- m_1 t_0}} {m_1 - m_2} \{ (-(\beta + \nu \beta + m_2)y_0 - \beta x_0 + c+ m_2 E_1) \nonumber\\
&& - (\omega E_2 - m_2 E_3 ) \cos \omega t_0 + (\omega E_3 + m_2 E_2 +f) \sin \omega t_0 \}  \\
C_2 =  &&\frac{e^ {- m_2 t_0}} {m_2 - m_1} \{ (-(\beta + \nu \beta + m_1)y_0 - \beta x_0 + c+ m_1 E_1) \nonumber\\
&& - (\omega E_2 - m_1 E_3) \cos \omega t_0 + (\omega E_3 + m_1 E_2 + f) \sin \omega t_0 \}
\end{eqnarray}
\label{ana5}
\end{subequations}

\subsection{$D_{+1}$ region}

In $D_{+1}$ region, $g(x) = bx + (a-b)$ and the state equations are written as
\begin{subequations}
\begin{eqnarray}
{dx \over dt }  &=&  y - b x - (a-b), \\
{dy \over dt }  &=&  -\beta y - \nu \beta y -\beta x + f \sin( \omega t) + c,
\end{eqnarray}
\label{ana6}
\end{subequations}
The stability of the fixed point $p_1 = (\frac{(\beta+\nu \beta) (b-a)+c}{\beta(1+b(\beta+\nu \beta))}, \frac{\beta(a-b)+bc}{b (\beta+\nu \beta)+\beta})$ is determined using the Jacobian matrix 
\begin{equation}
J_0 =
\begin{pmatrix}
-b &&& 1 \\
-\beta &&& - (\beta + \nu \beta) \\
\end{pmatrix},
\label{ana7}
\end{equation}
and $p_1$ is found to be a {\it{stable focus fixed point}}.
Equation (\ref{ana6}) can be written as a second-order differential equation in terms of the variable $y$ as
\begin{equation}
{\ddot y} + {A_2 \dot y} + B_2 y = \Delta_2 + b f \sin(\omega t) + f \omega \cos(\omega t),
\label{ana8}
\end{equation}
where $A_2 = \beta + \nu \beta + b$, $B_2 =\beta+ b (\beta+ \nu \beta)$ and $\Delta_2 = b c + \beta(a-b)$. For the given values of the system parameters, the roots of Eq. (\ref{ana8}) $m_{3,4} = u \pm iv$, where, $u=\frac{-A_2}{2}, v=\frac{\sqrt{4B_2 - {A_2}^2}}{2}$ are a pair of complex conjugates and the state variables are obtained as

\begin{subequations}
\begin{eqnarray}
y(t) &=& e^ {ut} (C_3 \cos vt + C_4 \sin vt) +E_4 +E_5 \sin(\omega t)+ E_6 \cos(\omega t),\\
x(t) &=& \frac{1}{\beta}(-(\beta + \nu \beta) y - \dot{y} + c + f \sin (\omega t)),
\end{eqnarray}
\label{ana9}
\end{subequations}
The constants $E_4, E_5, E_6$ are the same as $E_1, E_2, E_3$ of Eq. (\ref{ana5}) except the constants $\Delta_1, a$ are replaced with $\Delta_2, b$, respectively. The constants of the complementary function $C_3, C_4$ are
\begin{eqnarray}
C_3 =&&  \frac{e^ {- u t_0}} {v} \{ ((\beta+\nu \beta + u){y_0} + \beta {x_0} -c) \sin v t_0  + (v y_0 - v E_4 ) \cos v t_0 - u E_4 \sin v t_0 \nonumber \\
&& + ((\omega E_5 - u E_6) \sin v t_0 - v E_6 \cos v t_0) \cos \omega t_0 \nonumber \\
&& - ((\omega E_6 + u E_5+ f) \sin v t_0 + v E_5 \cos v t_0 ) \sin \omega t_0 \} \nonumber \\
C_4 =&&  \frac{e^ {- u t_0}} {v} \{ (c - (\beta+\nu \beta + u){y_0} - \beta {x_0}) \cos v t_0  + (v y_0 - v E_4 ) \sin v t_0 + u E_4 \cos v t_0 \nonumber \\
&& - ((\omega E_5 - u E_6) \cos v t_0 + v E_6 \sin v t_0) \cos \omega t_0 \nonumber \\
&& + ((\omega E_6 + u E_5+ f) \cos v t_0 - v E_5 \sin v t_0 ) \sin \omega t_0 \} \nonumber \nonumber
\end{eqnarray}

\subsection{$D_{-1}$ region}

In this region, $g(x) = bx - (a-b)$ and the state equations are written as
\begin{subequations}
\begin{eqnarray}
{dx \over dt }  &=&  y - b x + (a-b), \\
{dy \over dt }  &=&  -\beta y - \nu \beta y -\beta x + f sin( \omega t) + c,
\end{eqnarray}
\label{ana10}
\end{subequations}
The stability of the fixed point $p_2 = (\frac{(\beta+\nu \beta) (a-b)+c}{\beta(1+b(\beta+\nu \beta))}, \frac{\beta(b-a)+bc}{b (\beta+\nu \beta)+\beta})$ is obtained using Eq. (\ref{ana7}) and $p_2$ is found to be a {\it{stable focus fixed point}}.
The state equation written in terms of the variable $y$ is
\begin{equation}
{\ddot y} + {A_3 \dot y} + B_3 y = \Delta_3 + b f sin(\omega t) + f \omega cos(\omega t),
\label{ana11}
\end{equation}
where $A_3 = \beta + \nu \beta + b$, $B_3 =\beta+ b (\beta+ \nu \beta)$ and $\Delta_3 = b c - \beta(a-b)$. The roots of eq. (\ref{ana11}) are a pair of complex conjugates given as $m_{3,4} = u \pm iv$, where, $u=\frac{-A_3}{2}, v=\frac{\sqrt{4B_3 - {A_3}^2}}{2}$ and the state variables are obtained as
\begin{subequations}
\begin{eqnarray}
y(t) &=& e^ {ut} (C_5 \cos vt + C_6 \sin vt) +E_7 +E_8 \sin(\omega t)+ E_9 \cos(\omega t),\\
x(t) &=& \frac{1}{\beta}(-(\beta + \nu \beta) y - \dot{y} + c + f \sin (\omega t)),
\end{eqnarray}
\label{ana12}
\end{subequations}
The constants $E_7, E_8, E_9,C_5,C_6$ are the same as $E_4, E_5, E_6,C_3,C_4$ of $D_{+1}$ region except that $\Delta_2, a$ is replaced with $\Delta_3$. The analytical solutions are used to generate the trajectories in the $(x-y)$phase plane of the system by varying a control parameter of the system for the fixed values of other parameters. \\

\begin{figure}
\begin{center}
\includegraphics[scale=0.23]{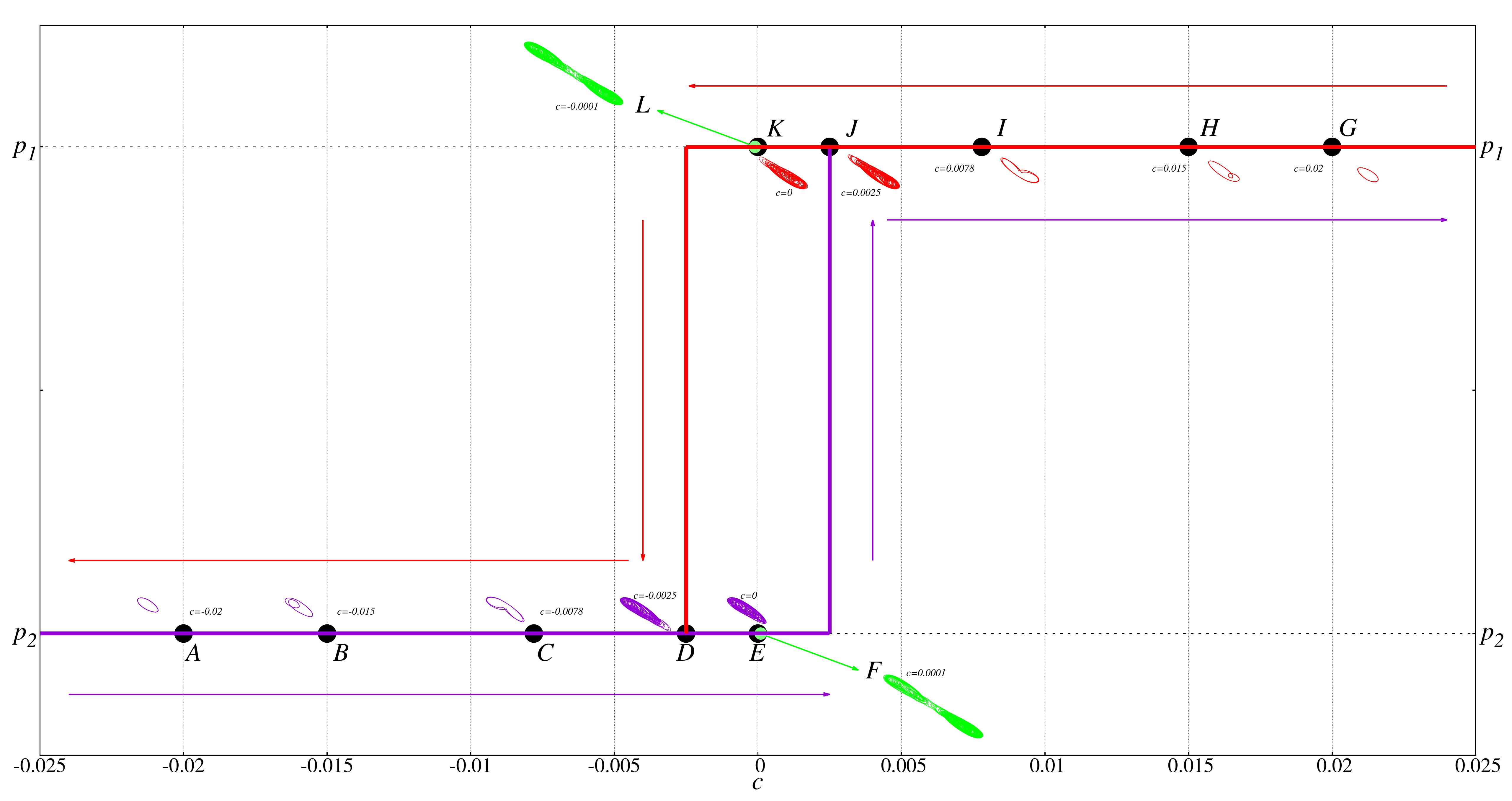}
\caption{Analytical results: Chaotic hysteresis observed in $MLC$ circuit through explicit analytical solutions. The system is initially operated in the one-band chaotic attractor state ($c=0$) existing in the left-half or right-half plane. The increase of the parameter $c$ from in the range $-0.02 \le c \le 0.02$ with the attractor fixed in the left-half plane ($c=0$) yields the dynamical path $ABCDEFJIHG$. The chaotic attractors in the $x-y$ plane is plotted adjacent to each of the dynamical path points. Conversely, the dynamical path $GHIJKLDCBA$ is observed when the attractor is initially set at the right-half plane ($c=0$) and for $c$ varying in the range $0.02 \ge c \ge -0.02$. The intersection of these dynamical paths reveals the existence of the chaotic hysteresis represented by the path $DEFJKL$. The jumping of the attractor from the left-halt to right-half plane and vice-versa occurs through the evolution of the double-band chaotic attractor state.}
\label{ana_fig1}
\end{center}
\end{figure}

The emergence of chaotic hysteresis in the system observed using the explicit analytical solutions briefed above is depicted in Fig. \ref{ana_fig1}. The MLC circuit is initially set for the one-band chaotic attractor state at $c=0$. The one-band chaotic attractor is so chosen such that it exists at the edge of the one-band chaotic state. The parameter values are kept at the values $a=-1.02, b=-0.55, \beta=1.0, \nu=0.015, f=0.1105, \omega=0.75$. Because of the dependence of the fixed points $p_0, p_1, p_2$ on the parameter $c$, a smaller shift in the position of the fixed points in the phase plane is observed while $c$ is varied. However, the stability of the fixed points remains unaffected as they are independent of $c$. Increasing $c$ value from $c=-0.02$ with the system operating initially ($c=0$) at the one-band chaotic attractor state around the fixed point $p_2$ in clockwise direction in the LHP for the initial condition $x_0, y_0 = -0.1, -0.1$ yields the dynamical path $ABCDEFJIHG$. The violet colored region indicates the dynamics of the existing in the LHP while the red colored corresponds to the dynamics in the RHP. The phase-portraits existing in the LHP corresponding to the points of the dynamical path is presented near each point. The different dynamical behaviors of the system at each of the points is briefed in Table 1. From Fig. \ref{ana_fig1} it is observed that the one-band chaotic attractor existing in the LHP (E) existing around the fixed point $p_2$ at $c=0$ jumps to the RHP as an one-band chaotic attractor (J) evolving around the fixed point $p_1$ in the clockwise direction at $c=0.0025$ through the evolution of the double-band chaotic attractor (F) at $c=0.0001$ and the system exhibits reverse period-doubling sequence with the attractors existing in the RHP represented by attractors and path in red. Similarly, decreasing $c$ from $c=0.02$ with the system operating initially ($c=0$) at the one-band chaotic state in the RHP around the fixed point $p_1$ yields the dynamical path $GHIJKLDCBA$. The transition of the one-band chaotic attractor state existing around $p_1$ in the RHP at $c=0$ to an attractor existing around $p_2$ in the LHP at the value $c=-0.0025$ (D) is observed through the evolution of a double-band chaotic attractor state at c=-0.0001 (L). The intersection of the paths $ABCDEFJIHG$ and $GHIJKLDCBA$ yields the region of chaotic hysteresis represented by $DEFJKL$ which is in confirmation with the dynamical region represented in the numerical simulation results shown in Fig. \ref{num_fig1}.  

\begin{table}
\begin{center}
\begin{tabular}{c c c c}
\hline
Dynamical Point & $c$ & Dynamical state & LHP/RHP  \\
\hline \\
A   		 			& 		-0.02	 	& period-1T LC 		& LHP   				 \\	
					&					&					&					\\	    		
B   					& 		-0.015 		& period-1T LC		& LHP     			     \\
					&					&					&					\\	
C   		 			& 		-0.0078	 	& period-4T LC 		& LHP    				 \\	
					&					&					& 								\\	    		
D   					& 		-0.0025 		& One-band chaos	& LHP     			     \\
					&					&					&								\\
E   					& 		0.0 			& One-band chaos	& LHP     			     \\
					&					&					&								\\
F   					& 		0.0001 		& Double-band chaos	& Both  			     \\
					&					&					&								\\
G   		 			& 		0.02	 	& period-1T LC    	& RHP 				 \\	
					&					&					&								\\	    		
H   					& 		0.015 		& period-1T LC    	& RHP 		     \\
					&					&					&			\\	
I   		 			& 		0.0078	 	& period-4T LC    	& RHP 			 \\	
					&					&					&			\\	    		
J   					& 		0.0025 		& One-band chaos 	& RHP 		     \\
					&					&					&			\\
K   					& 		0.0 			& One-band chaos    	& RHP 		     \\
					&					&					&			\\
L  					& 		-0.0001 		& Double-band chaos 	& Both 		     \\
					&					&					&			\\
\hline
\end{tabular}
\caption{Dynamics of the MLC system at the dynamical points specified in Fig. \ref{ana_fig1}. LC-Limit Cycle.}
\label{tab:1}
\end{center}
\end{table}         

\begin{figure}
\begin{center}
\includegraphics[scale=0.3]{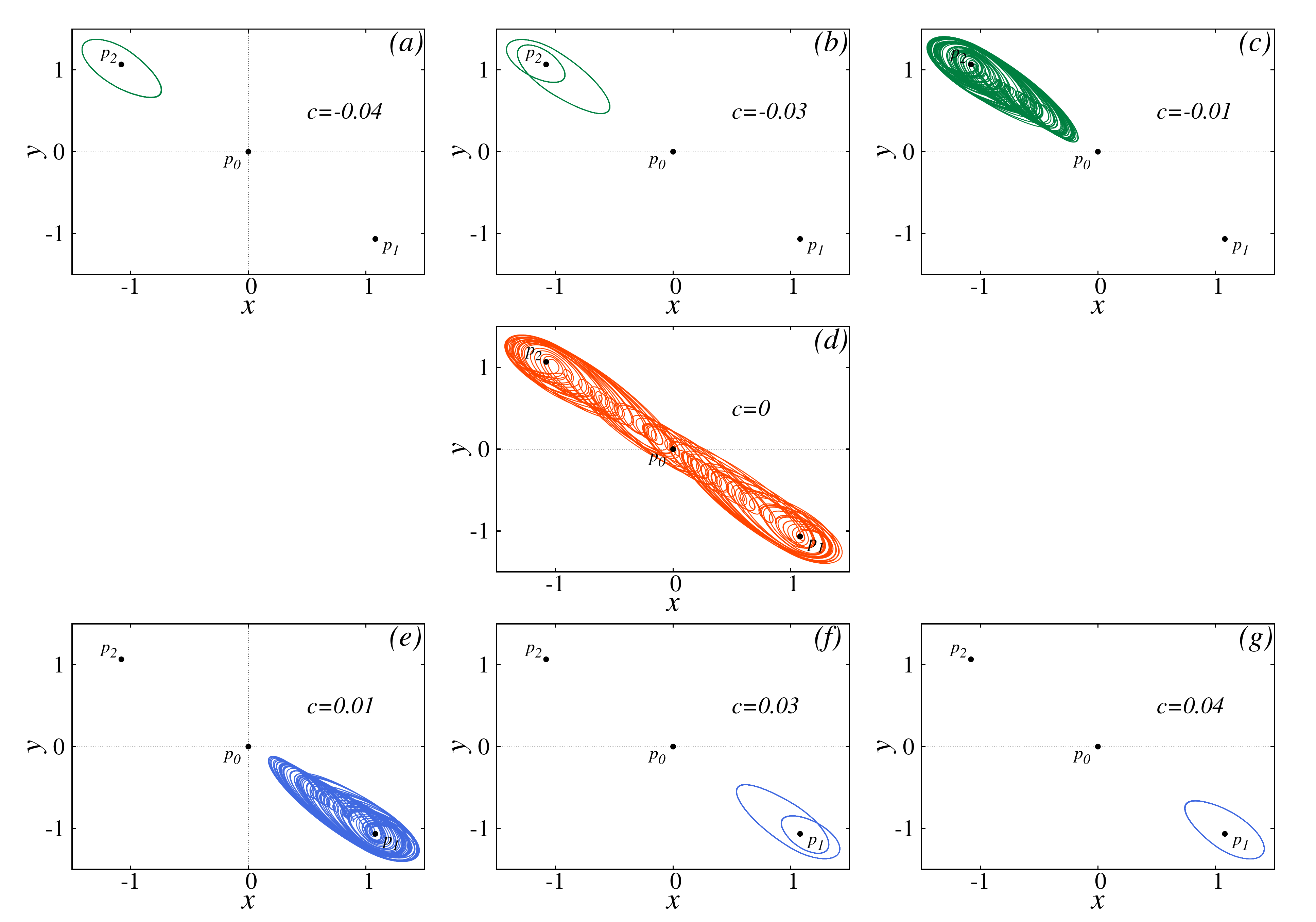}
\caption{Analytical results indicating the phenomena of chaotic hysteresis in MLC circuit for the offset voltage factor $c$ as the control parameter. (a) Period-T attractor for $c=-0.04$, (b) Period-2T attractor for $c=-0.03$ and (c) one-and chaotic attractor for $c=-0.01$ obtained in the LHP around the fixed point $p_1$; (d) double-band chaotic attractor for $c=0$; (e) one-and chaotic attractor for $c=0.01$, (f) Period-2T attractor for $c=0.03$ and (g) Period-T attractor for $c=0.04$ obtained in the RHP around the fixed point $p_2$.}
\label{ana_fig2}
\end{center}
\end{figure}

\begin{figure}
\begin{center}
\includegraphics[scale=0.23]{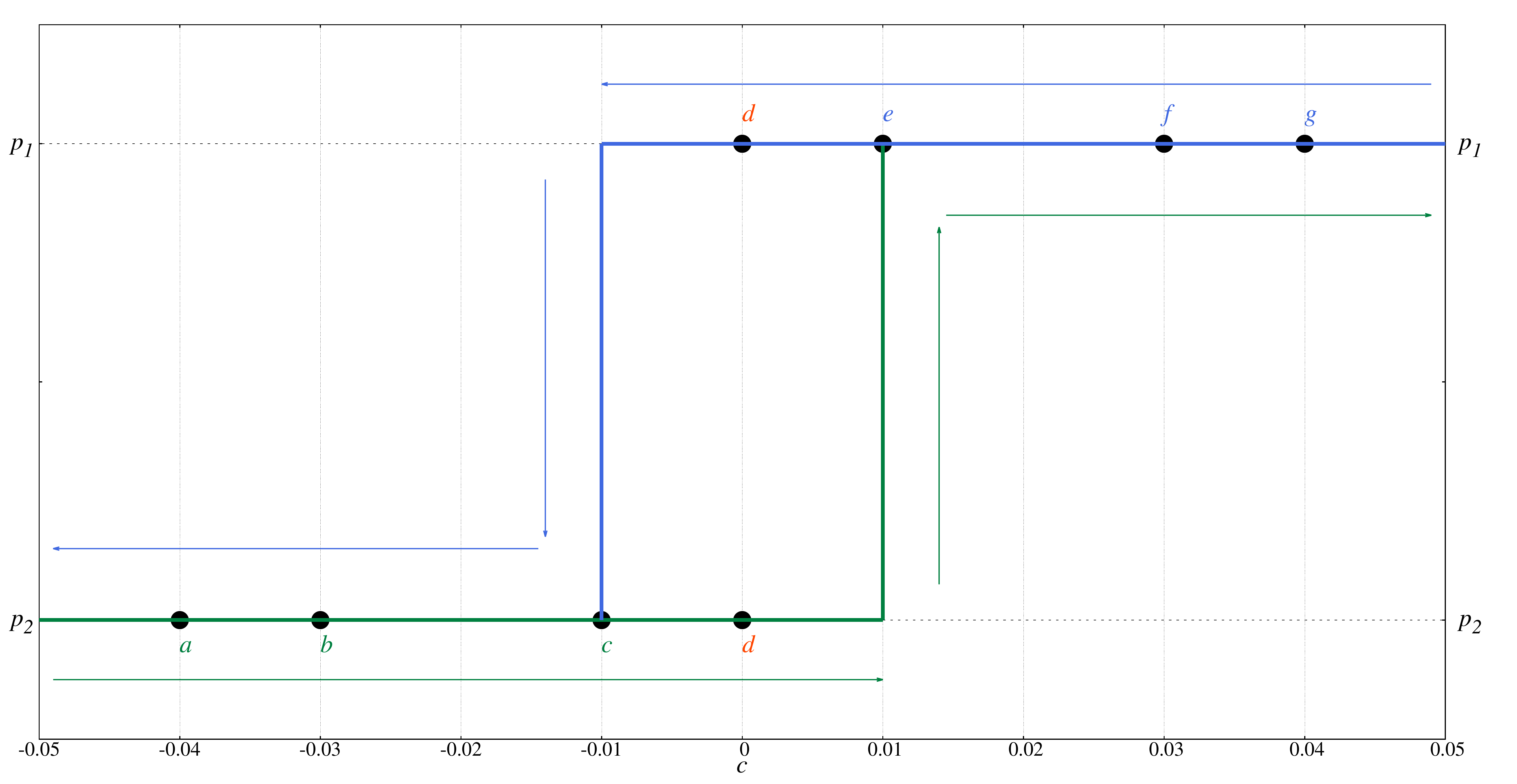}
\caption{Analytical results indicating chaotic hysteresis in the MLC circuit with the variation of DC offset voltage factor $c$. The labels $a-g$ in the figure corresponds to the phase-portraits in Fig. \ref{ana_fig2}(a)-\ref{ana_fig2}(g).}
\label{ana_fig3}
\end{center}
\end{figure}

The chaotic hysteresis phenomenon observed in the $MLC$ circuit with the system initially operating in the double-band chaotic attractor state is depicted in Fig. \ref{ana_fig2}. The circuit is operated at the double-band chaotic state for $c=0$ and the other parameters are set as $a=-1.02, b=-0.55, \beta=1.0, \nu=0.015, f=0.14, \omega=0.75$. The phenomenon of chaotic hysteresis is observed as the parameter $c$ is increased in the range $-0.04 \le c \le 0.04$. The system exhibiting a period-T limit cycle attractor at $c=-0.04$ about the fixed point $p_2$ evolves into a period-2T limit cycle at $c=-0.03$ before settling down into an one-band chaotic attractor at $c=-0.01$ as shown in Figs. \ref{ana_fig2}(a), \ref{ana_fig2}(b) and \ref{ana_fig2}(c), respectively. Increasing the value of $c$ results in a double-band chaotic attractor at $c=0$ as shown in Fig. \ref{ana_fig2}(d). Further increase in $c$ results in a one-band chaotic attractor evolving around the fixed point $p_2$ in the RHP at $c=0.01$ leading to a period-2T attractor at $c=0.03$ as shown in Figs. \ref{ana_fig2}(e) and \ref{ana_fig2}(f), respectively. The system finally settles down to a period-T limit cycle attractor at $c=0.04$ as observed n Fig. \ref{ana_fig2}(g). The chaotic hysteresis behavior is also observed when the parameter is decreased from $0.04$ to $-0.04$. In either case, the hopping of the attractors around the fixed points $p_1$ to $p_2$ and {\it{vice versa}} about the double-band chaotic attractor state for equal and opposite values of $c$ indicating the phenomenon of chaotic hysteresis is observed. The evolution of chaotic hysteresis around the double-band chaotic attractor state is depicted in Fig. \ref{ana_fig3}. The dynamical points $a,b,c,d,e,f,g$ given in Fig. \ref{ana_fig3} indicates the dynamical state of the system as given in the phase portraits of Fig. \ref{ana_fig2}(a)-\ref{ana_fig2}(g), respectively. Chaotic hysteresis occurs in the system through hopping of the one-band chaotic attractors across the planes and by encompassing the double-band attractor state.

\section{Multistability analysis}
\begin{figure*}[!htpb]
	\centering
	\subfloat{\includegraphics[width=6.5cm, height =5cm]{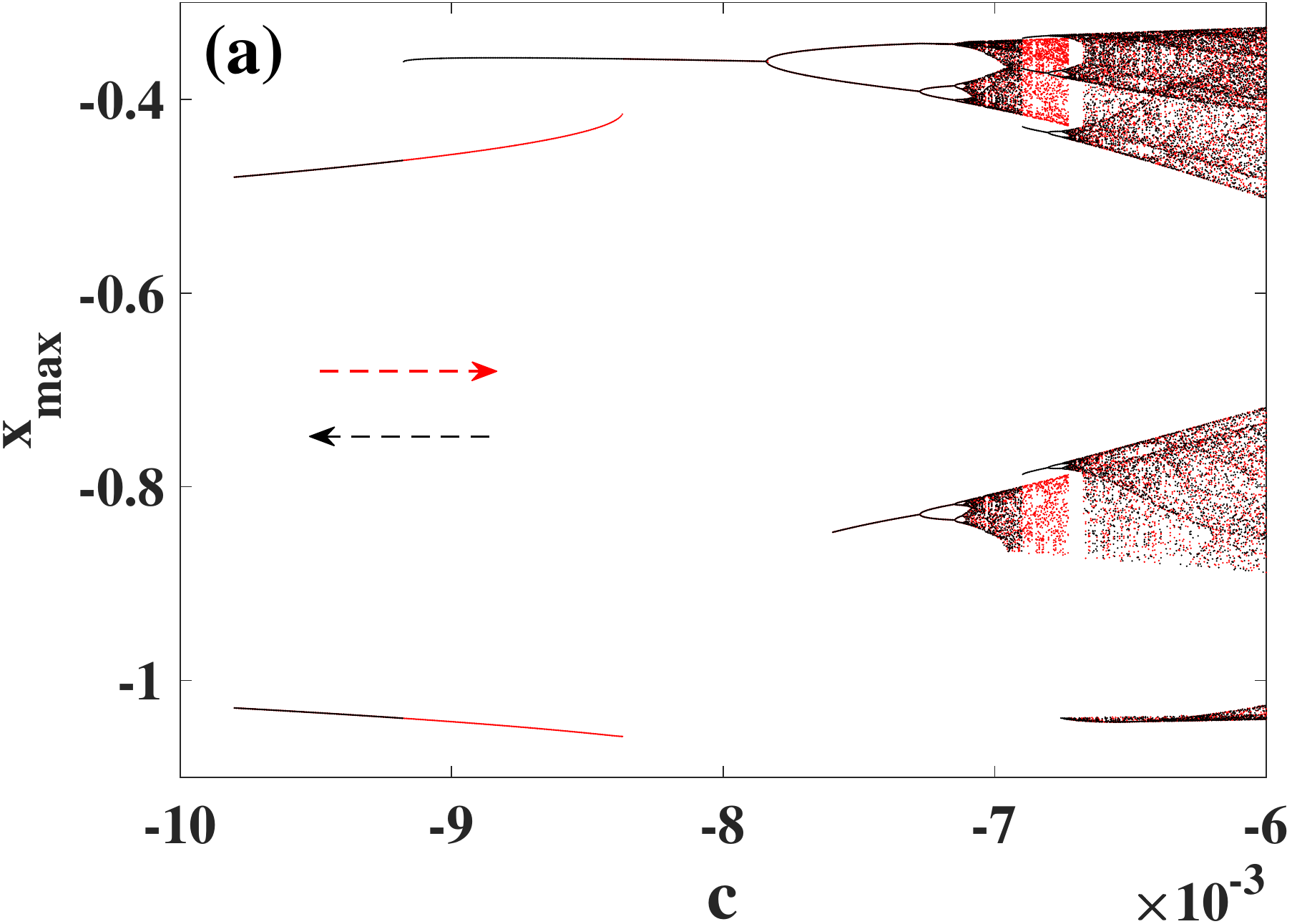}}
	\subfloat{\includegraphics[width=6.5cm, height =5cm]{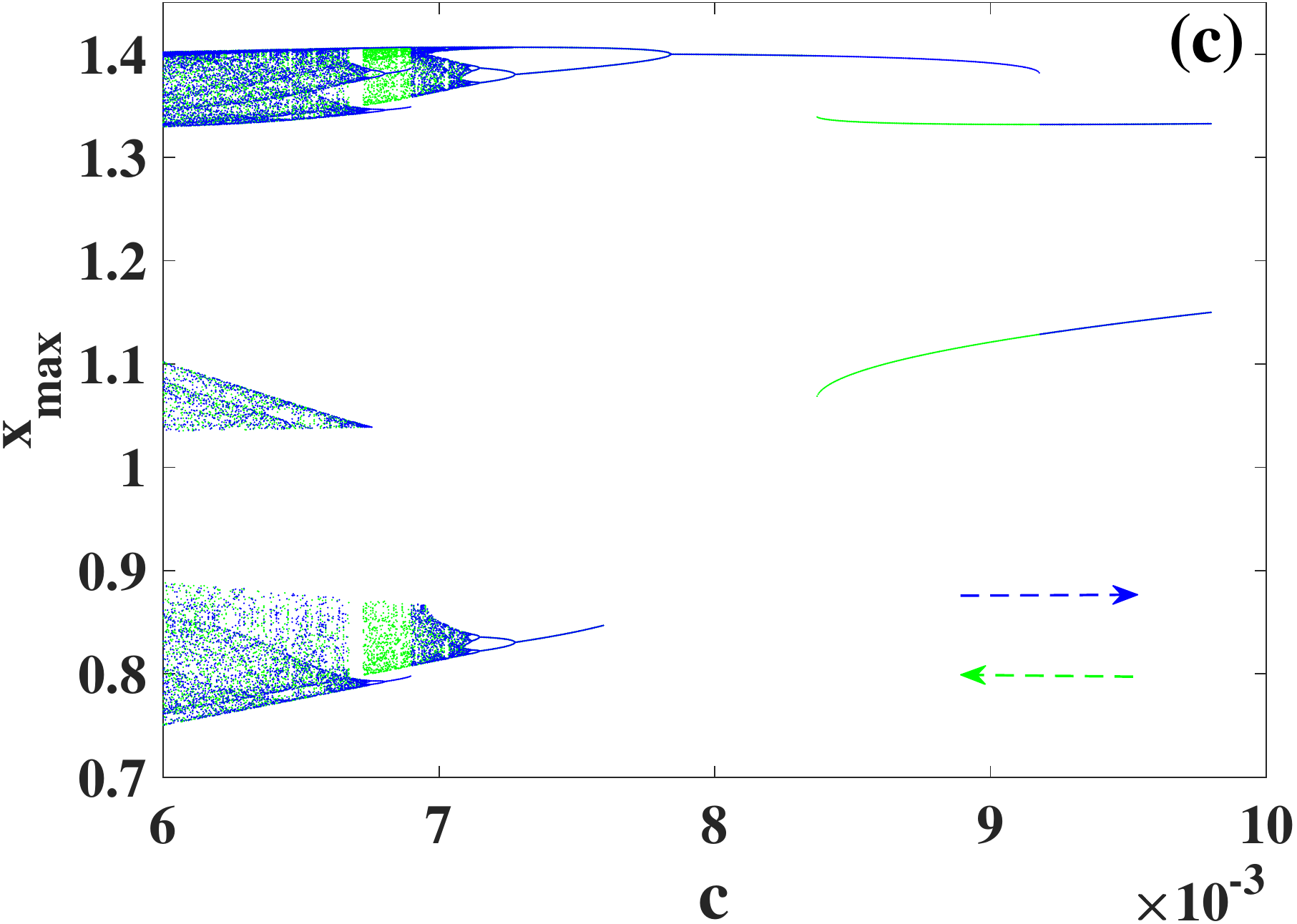}}\\
	\subfloat{\includegraphics[width=6.5cm, height =5cm]{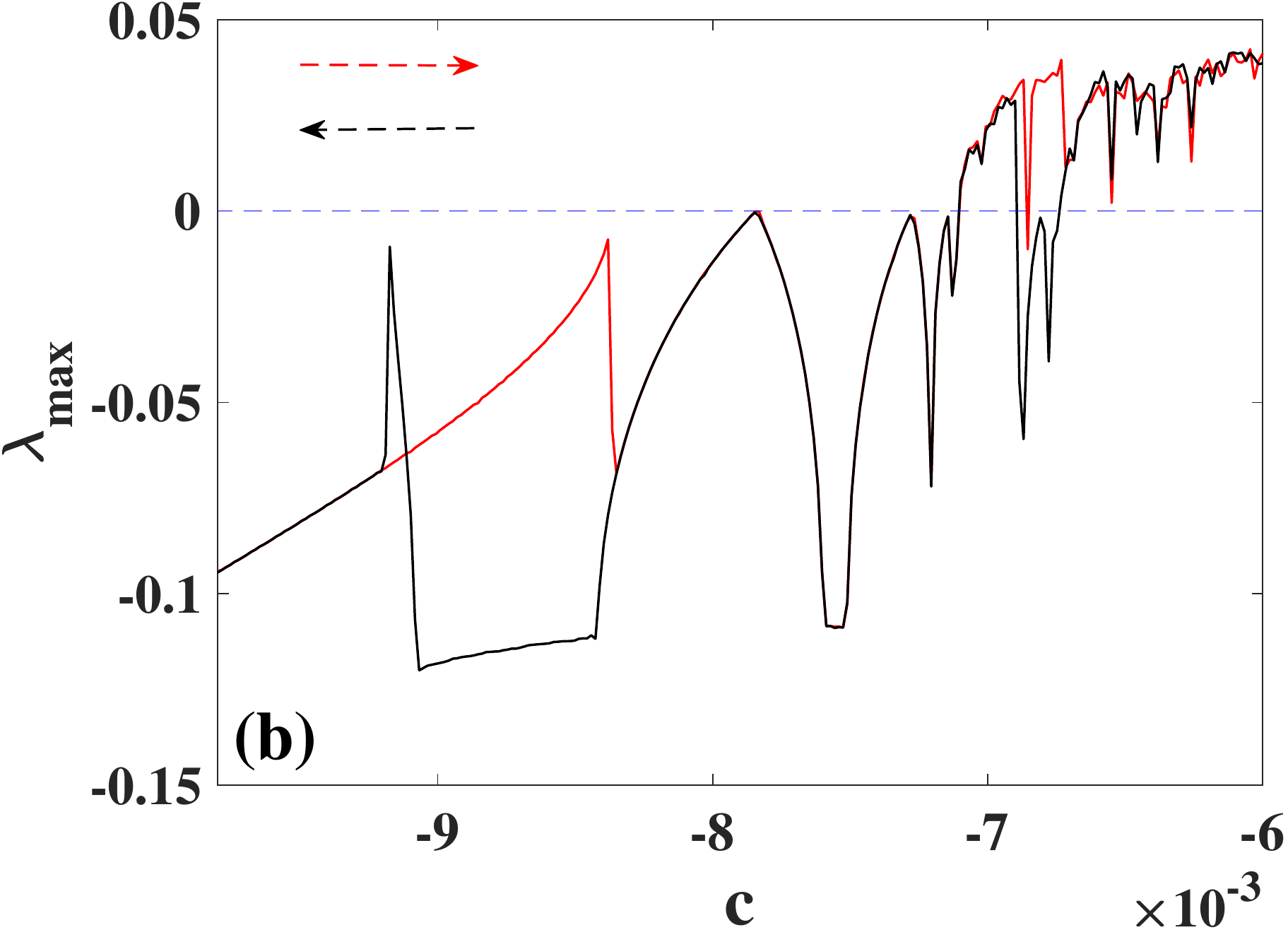}}
	\subfloat{\includegraphics[width=6.5cm, height =5cm]{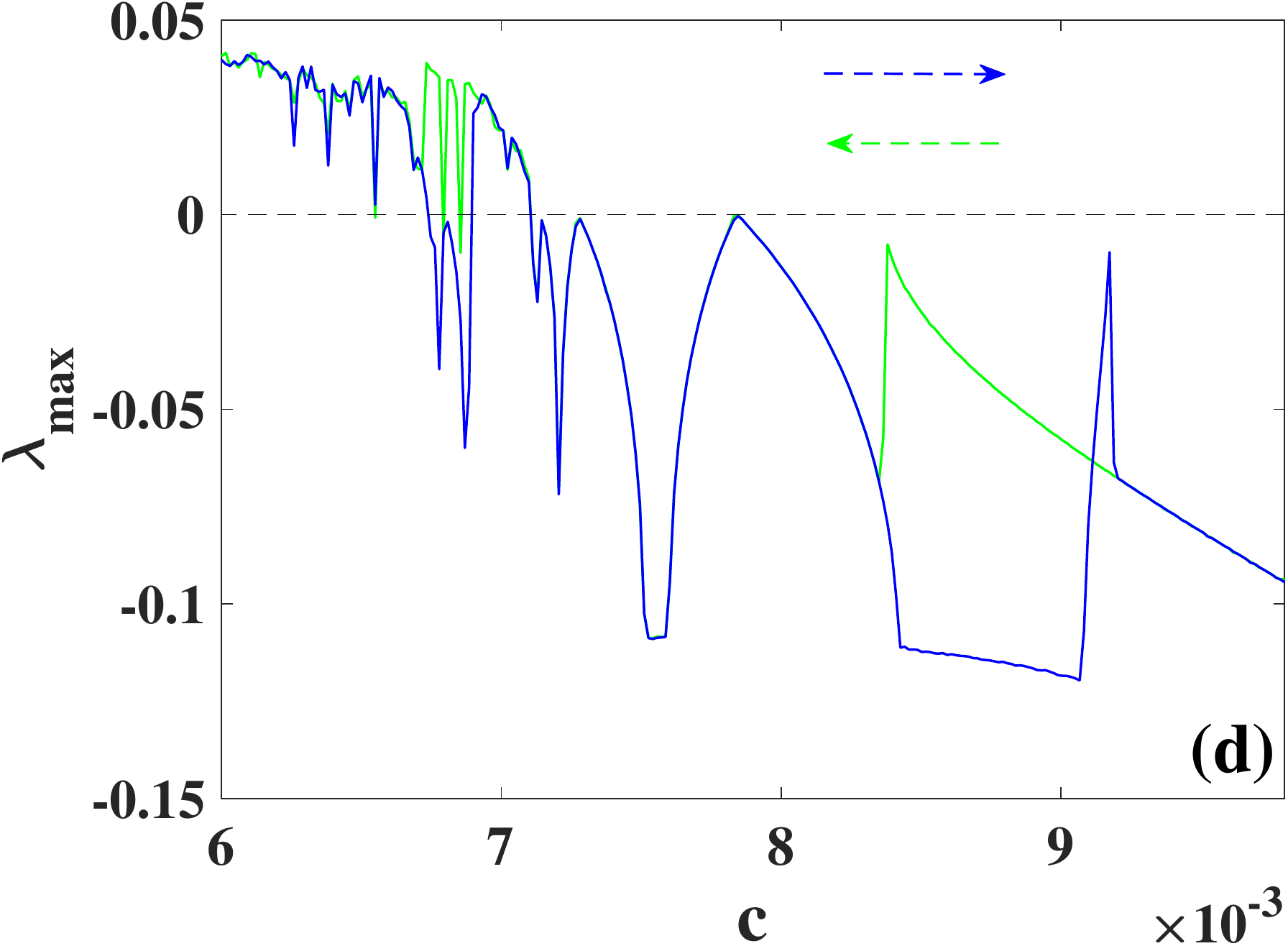}}	
	\caption{Magnified region of Fig.\ref{num_fig1} and \ref{num_fig2} showing the coexistence of different dynamics in system (\ref{num_1}). (a) Bifurcation diagrams (red and black colors) and (b) graphs of maximum Lyapunov exponents for $c\in [-9.8,-6]\times 10^{-3}$. (c) Bifurcation diagrams (in blue and green colors) and (d) graphs of maximum Lyapunov exponents for $c\in [6,9.8]\times 10^{-3}$. Various regions with the coexistence of two qualitative and quantitative dynamics are clearly visible. (color online)}
	\label{fig:Coexistence}
\end{figure*}

Figures \ref{num_fig1} and \ref{num_fig2} illustrate the bifurcation and maximum Lyapunov exponent as functions of the control parameter $c$. It is observed that small periodic windows appear around $c=-0.007$ and $c=0.007$. To provide a closer view of these regions, we have included magnified versions of the bifurcation diagrams and the maximum Lyapunov exponent plot in Fig. \ref{fig:Coexistence}. In this figure, we have plotted the data for $c$ ranging from -0.025 to 0.025 and backwards, using different colors to represent the two sets of data superimposed on top of each other. The results show that tiny regions of multistability exist in the bifurcation space (Fig. \ref{fig:Coexistence}(a),\ref{fig:Coexistence}(c)) and maximum Lyapunov exponent plot (Fig. \ref{fig:Coexistence}(b),\ref{fig:Coexistence}(d)), indicating the coexistence of periodic and chaotic behavior. To investigate the existence of these multistability regions in equation (\ref{num_1}), we have used an offset boost in the $y$ variable.



\subsection*{Offset boosted MLC circuit}
\label{sec:2-2}

	Offset boosting is an easy and reliable method to identify any coexisting dynamics in the phase space \cite{Sivaganesh2023,li2017diagnosing,fozin2022coexistence}. Offset boosted systems are obtained by introducing a constant vector into the state variables of an investigated system. This causes a shift in the opposite direction of the sign of the introduced offset boosting variable in the state space with no alteration on the system dynamics.  
	System (\ref{num_1}) is offset boosted  over the variable $y$ with $m$ as the boosting controller as depicted in Eq(\ref{Eq:2D-2}).

\begin{equation}
\label{Eq:2D-2}
\left\{ \begin{array}{lcl}
\vspace{0.1cm}
\displaystyle  \dot{x} & = & \displaystyle y + m  - g(x) \\\vspace{0.1cm}
\displaystyle \dot{y} & = & \displaystyle -\sigma (y + m)  - \beta x + f \sin(\omega t)  + c
\end{array} \right.
\end{equation}

\begin{figure*}[!htpb]
	\centering
	\subfloat[][]{\includegraphics[width=6cm, height =5cm]{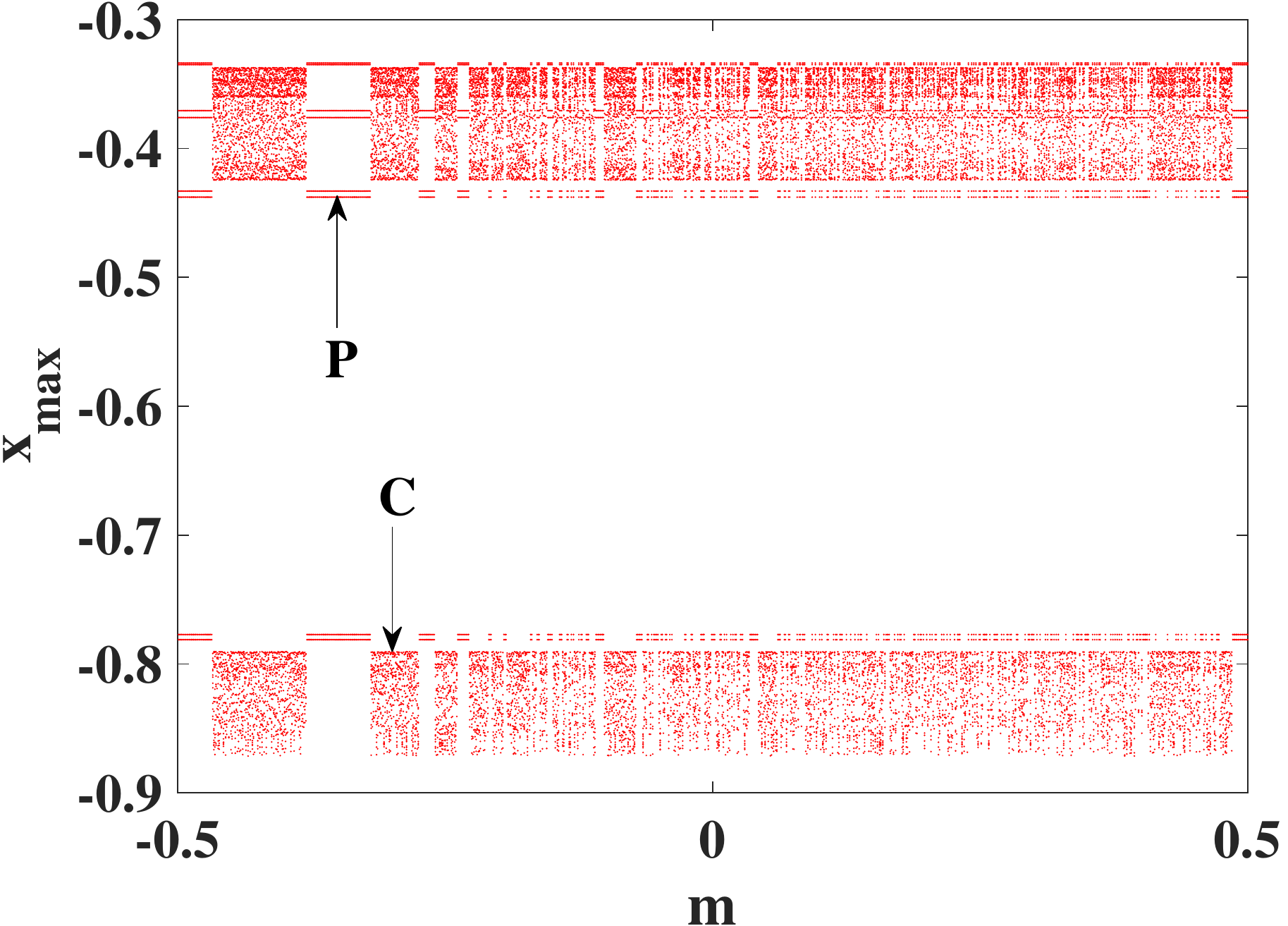}}
	\subfloat[][]{\includegraphics[width=6.5cm, height =5.25cm]{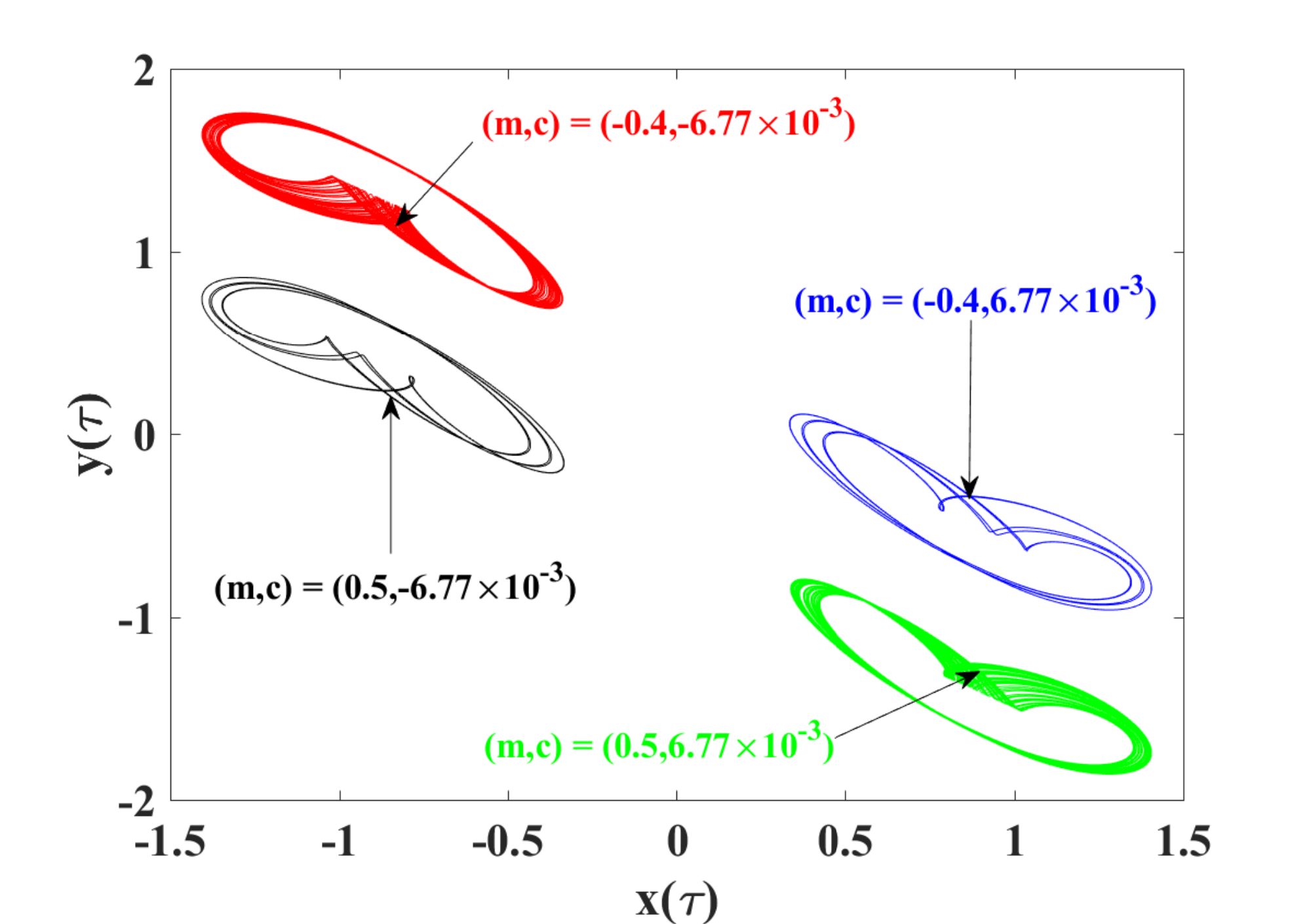}}	
	\caption{(a) Bifurcation diagram of system (\ref{Eq:2D-2}) versus the offset boosting parameter $m \in [-0.5,0.5]$ for $c = -6.77 \times 10^{-3}$ showing the periodic/regular oscillations labeled with $P$ and chaotic dynamics denoted by $C$. (b) Phase portraits showing the coexisting attractors in the plane $(x,y)$ for various sets of $(m,c)$ (i.e,the offset boosting parameter and bifurcation term respectively).  The color code of the coexisting attractors are receptively those of Fig.\ref{fig:Coexistence} for $c=\mp 6.77 \times 10^{-3}$ (color online).}
	\label{fig:portrait_coexistence}
\end{figure*}

When $c$ is fixed at $-6.77\times 10^{-3}$ and the boosting controller $m$ is varied within $[-0.5,0.5]$, the bifurcation diagram in Fig.\ref{fig:portrait_coexistence}(a) illustrates two distinct dynamics, one regular/periodic and the other chaotic. Two discrete offset boosting controller values, $m=-0.4$ and $m=0.5$, yield captured phase portraits in red and black colors, respectively, as shown in Fig.\ref{fig:portrait_coexistence}(b). The coexisting phase portraits in green (chaotic) and blue (periodic) colors are obtained for the respective values of $(m,c) = (0.5,6.77\times 10^{3})$ and $(m,c) = (-0.4, 6.77\times 10^{3})$, as shown in Fig.\ref{fig:portrait_coexistence}(b).

\section{Conclusion}
\label{sec_con}

Chaotic hysteresis is observed in {\it{Murali-Lakshmanan-Chua}} (MLC) circuit when a DC offset voltage is introduced and varied.  The MLC circuit initially is set to operate in the one-band chaotic state, in the absence of an offset voltage parameter.  In this one-band chaotic state, the chaotic motion moves from left-half plane of the phase space to the right-half plane of the phase space through the evolution of a double-band chaotic attractor for a certain value of the offset voltage as the offset voltage parameter is increased. Further, when the offset voltage parameter is decreased, chaotic motion moves from right-half plane to the left-half plane again through the evolution of a double-band chaotic attractor but for a different value of an offset voltage, producing a hysteresis loop.  The existence of hysteresis over a certain region of the offset voltage parameter encompassing the initial state is observed and reported. The circuit can be set to operate at different initial dynamical states to study the emergence of chaotic hysteresis through various routes.  The phenomenon of chaotic hysteresis observed by electronic circuit experiments is validated through numerical and analytical results.  Multistable regions are also observed in the MLC circuit with DC offset voltage and studied using offset boosting.

\section*{Acknowledgement}

The authors are grateful to Dr.K. Murali, Professor, Department of Physics, Anna University, Chennai, India for his valuable suggestions and discussions.

\bibliographystyle{unsrt}
\bibliography{mybibfile_1}

\end{document}